# Depth-Based Vector Median Absolute Deviation Moments for Robust Multivariate Shape Analysis

**Elsayed A. H. Elamir**

Department of Management and Marketing, College of Business Administration,
Kingdom of Bahrain
Email: shabib@uob.edu.bh

**Abstract**

Classical multivariate shape analysis relies on covariance standardized moments, such as Mardia's skewness and kurtosis, which are sensitive to outliers and require finite moments. This paper introduces vector median absolute deviation (VMedAD) moments for robust multivariate shape analysis. The proposed framework replaces moment aggregation and covariance standardization with median-based centre–outward contrasts defined through data depth, yielding affine-equivariance, moment-free vector moments. VMedAD moments provide direction preserving measures of multivariate skewness and directional peripheral dominance, separating central structure from tail driven behaviour. Consistency, breakdown properties, and affine equivariance are established, and simulation and real dataset examples demonstrate improved robustness and geometric interpretability over classical and projection-based methods.

# 1 Introduction

Classical measures of multivariate skewness and kurtosis are rooted in covariance-standardized moments, most notably Mardia's (1970, 1974) extensions of univariate third and fourth order moments. These measures have played a central role in assessing departures from elliptical symmetry and multivariate normality and remain widely used in theory and practice (Kent & Bibby, 1979). Let $\boldsymbol{X} \in \mathbb{R}^d$ be a random vector with mean $\boldsymbol{\mu}$ and positive-definite covariance matrix $\boldsymbol{\Sigma}$. Mardia defined the population multivariate skewness and kurtosis as

$$\beta_{1,d} = \mathrm{E}[((\boldsymbol{X} - \boldsymbol{\mu})^\top \boldsymbol{\Sigma}^{-1} (\boldsymbol{X} - \boldsymbol{\mu}))^3], \text{ and } \beta_{2,d} = \mathrm{E}\left[((\boldsymbol{X} - \boldsymbol{\mu})^\top \boldsymbol{\Sigma}^{-1} (\boldsymbol{X} - \boldsymbol{\mu}))^2\right].$$

These measures are affine invariant and reduce, under multivariate normality, to constants depending only on the dimension $d$. However, their reliance on finite second and higher-order moments, together with their dependence on covariance matrices, renders them sensitive to outliers and ill-defined under heavy tailed distributions.

Recent work on multivariate skewness and kurtosis has considerably expanded the classical framework, primarily through refined moment-based formulations and parametric generalizations. For example, (Zuo et al., 2026) provide a comprehensive synthesis of multivariate skewness measures and Mardia's kurtosis within skew-elliptical families, deriving explicit expressions under canonical representations. This line of work offers valuable analytical insight and facilitates systematic comparison across parametric models. However, these measures remain fundamentally tied to third and fourth order moments, and covariance dependent (Zuo et al., 2026; Loperfido, 2024; Chowdhury et al., 2024). Although some recent proposals improve numerical behaviour or introduce alternative constructions, these measures typically aggregate information globally and reduce inherently directional aspects of multivariate asymmetry to scalar summaries (Baillien et al., 2024; Kim & Kim, 2025).

Several robust extensions of classical moments have been proposed, replacing means and covariances with medians and robust scatter estimators, or relying on projection-based constructions (Malkovich & Afifi, 1973; Oja, 2010). While these approaches improve resistance to outliers and heavy tails, they remain conceptually tied to moment aggregation of standardized residuals.

The vector median absolute deviation (VMedAD) moments proposed in this paper address these gaps by constructing multivariate shape measures directly from depth defined shells and median-based contrasts. This yields vector valued measures of multivariate skewness and directional peripheral dominance that remain well defined under heavy tailed distributions. Moreover, it separates central structure from peripheral effects and provide a clear geometric interpretation of multivariate asymmetry. In this sense, the proposed framework extends the role of data depth from symmetry diagnostics to a full moment analogue theory of multivariate shape. Finally, VMedAD improves robustness without collapsing interpretation and provides explicit, vector-valued, and scale-free shape descriptors whose directions and magnitudes are interpretable. We emphasize that the proposed methodology is not intended as a replacement for classical covariance-based analysis, but rather as a full robust analogue and completion that remains well defined under heavy-tailed distributions and in the presence of outliers.

This study organized as follows, Section 2 presents the univariate MedAD. Section 2 proposes the extension to multivariate setting in term of population and sample moments and establishes their theoretical properties including consistency, breakdown point and affine equivariance. Section 4 presents simulation data illustrating robustness and geometric interpretability, moments for normal and $t$ distributions and application to real dataset. Section 5 concludes with a conclusion and directions for future research.

## 2 MedAD moments

Let $X_1, \ldots, X_n$ be an independent and identically distributed random sample drawn from a continuous population with distribution function $F_X(\cdot)$, where $0 < F_X(x) < 1$, density $f_X(\cdot)$ with $f_X(x) \geq 0$, and quantile function $Q(F)$. Let the population mean be $\mu = E(X)$, the population median be $M = \text{Med}(X)$, and the standard deviation be $\sigma = \sqrt{E[(X-\mu)^2]}$. Denote by $\mathbf{I}_{(i \leq k)}$ the indicator function, which equals 1 if $i \leq k$ and 0 otherwise. The corresponding order statistics of the sample are written as $X_{(1)}, \ldots, X_{(n)}$.

For integers $b \geq 0$, Elamir (2026) defines the $(b+1)$-th median absolute deviation moment by

$$\Phi_{b+1} = \begin{cases} M, & b = 0 \\ \text{Med}|X - M|, & b = 1 \\ \sum_{a=0}^{b-1} (-1)^{a+1} \text{Med}\big(|X - M|\ \mathbf{I}_{X \in Q(u,v]}\big), & \text{For } b \geq 2, \end{cases}$$

Where $Q(u, v] = Q_X(u = a/b, v = (a+1)/b]$ denotes the quantile slice corresponding to the interval $(u, v]$ of the cumulative distribution function of $X$, and $\mathbf{I}(\cdot)$ is the indicator function. The second MedAD moment,

$$\Phi_2 = \text{Med}(|X - M|),$$

serves as a robust scale functional and is the univariate median analogue of the standard deviation (Rousseeuw, 1987; Rousseeuw & Croux, 1993; Falk, 1998).

To remove scale dependence, the standardized MedAD moments are defined for $b \geq 2$ as

$$\Psi_{b+1} = \frac{\Phi_{b+1}}{\Phi_2} = \frac{\sum_{a=0}^{b-1}(-1)^{a+1}\text{Med}\big(|X - M|\ \mathbf{I}_{X \in Q(u,v]}\big)}{\text{Med}|X - M|}, \quad b \geq 2$$

These standardized quantities are dimensionless indices of distributional shape and are therefore directly comparable across datasets and measurement units.

For example, when $b = 2$, the standardized third MedAD moment

$$\Psi_3 = \frac{\Phi_3}{\Phi_2} = \frac{-\text{Med}\big(|X - M|\ I_{X \in Q(0,1/2]}\big) + \text{Med}\big(|X - M|\ I_{X \in Q(1/2,1]}\big)}{\text{Med}|X - M|}$$

provides a robust measure of skewness based on median-centred absolute deviations. Similarly, for $b = 3$, the standardized fourth MedAD moment

$$\Psi_4 = \frac{\Phi_4}{\Phi_2} = \frac{-\text{Med}\big(|X - M|\ I_{X \in Q(0,1/3]}\big) + \text{Med}\big(|X - M|\ I_{X \in Q(1/3,2/3]}\big) - \text{Med}\big(|X - M|\ I_{X \in Q(2/3,1\}}\big)}{\text{Med}|X - M|}$$

captures the degree of peripheral concentration around the centre of the distribution.

## 3 Vector median absolute deviation moments (VMedAD)

### 3.1 Population VMedAD

We extend the univariate framework introduced by Elamir (2026) to the multivariate setting, drawing inspiration from Falk (1997), who defined the comedian (covariance based on medians) between two random variables $X$ and $Y$ as

$$\text{COM}(X, Y) = \text{Med}[(X - \text{Med}(X))(Y - \text{Med}(Y))].$$

In the univariate case, taking $Y = X$ yields

$$\Phi_2^2 = \text{COM}(X, X) = \text{Med}(X - \text{Med}(X))^2,$$

which serves as the median variance analogue of the classical variance $\sigma^2 = \text{Var}(X) = \text{Cov}(X, X)$. Accordingly, $\Phi_2 = \{\text{Med}[(X - \text{Med}(X))^2]\}^{1/2}$, which we refer to as the median standard deviation.

Now let $X \in \mathbb{R}^d$ and denote by $\mathbf{M}$ a multivariate median of $X$ (e.g., spatial or component-wise). Define the centered random vector

$$\mathbf{U} = X - \mathbf{M}.$$

The median covariance matrix is then defined as

$$C_{\text{Med}} = \text{Med}(\mathbf{U}\mathbf{U}^T).$$

Equivalently, the $(i, j)$th entry is given by

$$(C_{\text{Med}})_{ij} = \text{Med}[(X_i - \mathbf{M}_i)(X_j - \mathbf{M}_j)].$$

The matrix $C_{\text{Med}}$ is the MedAD analogue of the classical covariance matrix. In this multivariate setting, the scalar second MedAD moment becomes

$$\Phi_2 = \sqrt{\text{Med}(\|X - \mathbf{M}\|^2)},$$

or, equivalently,

$$\Phi_2^2 = \text{tr}(C_{\text{Med}}).$$

This quantity represents the robust total median variance, providing a median-based measure of overall dispersion that remains well defined under heavy-tailed distributions and in the presence of outliers.

The VMedAD moments are proposed as follows. For an integer $b \geq 0$. The VMedAD moments can be proposed as

$$\Phi_{b+1} = \begin{cases} \mathbf{M}, & b = 0 \\ \text{Med}(X - \mathbf{M}), & b = 1 \\ \sum_{a=0}^{b-1}(-1)^{a+1}\text{Med}((X - \mathbf{M})\mathbf{I}(X \in S_a)), & b \geq 2, \end{cases}$$

Here $\mathbf{M}$ is median vector and $S_a$ represents the depth shell corresponding to the interval $(a/b, (a+1)/b]$ of the depth distribution where the depth distribution is divided into $b$ equal probability shells

$$S_a = \{\mathbf{x}: q_{a/b} \leq D(\mathbf{x}) < q_{(a+1)/b}, a = 0, 1, \ldots, b - 1\}$$

and $q_{a/b}$ is the $a/b$-quantile of the depth distribution. These shells induce a center-outward depth slicing, providing a multivariate analogue of univariate quantile intervals and forming the geometric foundation of the VMedAD scaling mechanism.

Where there is no natural ordering of observations in multivariate setting analogous to univariate quantiles, depth transforms unordered multivariate observations into a meaningful centre–outward ranking. Several depth notions exist; each emphasizes different aspects of multivariate structure, such as, Tukey half space, Mahalanobis, projection and spatial (Liu et al., 1999). Since all MedAD moments $\Phi_k$ are defined through median deviations within depth shells, making the choice of depth function crucial. Our choice of depth will be spatial depth

$$D_{sp}(x, F) = 1 - \left\|E\left(\frac{X - x}{\|X - x\|}\right)\right\|,$$

it is intrinsic and fully multivariate no projections and affine equivariant, (Mosler, 2024 and Serfling, 2004). This yields a multivariate analogue of univariate quantile slicing, but in a geometry aware way. this depth shells work as following.

- Compute depth $D(x_i)$ for data and order observations by depth,
- Split ordered depths into equal probability blocks and each block is a depth shell.
- Use this depth shell to obtain quantile and $S_a$.

The location functional is defined as
$$\boldsymbol{\Phi}_1 = \mathbf{M},$$
where $\mathbf{M}$ denotes a multivariate median and serves as a robust analogue of the classical mean vector $\boldsymbol{\mu}$. The second vector MedAD moment is defined as
$$\boldsymbol{\Phi}_2 = \text{Med}(\mathbf{X} - \mathbf{M}),$$
and quantifies directional imbalance in the multivariate distribution. This vector plays a role analogous to the classical moment $E(\mathbf{X} - \boldsymbol{\mu}) = \mathbf{0}$, while remaining well defined under heavy-tailed distributions and in the presence of outliers.

The VMedAD norm of the second vector moment services as a scalar measure of dispersion and is defined as
$$\Phi_2^{Med} = \|\boldsymbol{\Phi}_2\|_{Med} = \sqrt{\text{Med}(\|\mathbf{X} - \mathbf{M}\|^2)},$$
This quantity serves as a robust measure of overall dispersion and is referred to as the VMedAD scale. This is a median-based analogue of the Euclidean norm induced by the covariance matrix, replacing second moments by medians of squared radial distances which provides a robust analogue to $\sqrt{\text{tr}(\boldsymbol{\Sigma})}$, representing overall variability. As such, it yields a robust measure of global dispersion. The associated median covariance matrix is defined by
$$\mathbf{C}_{Med} = \text{Med}((\mathbf{X} - \mathbf{M})(\mathbf{X} - \mathbf{M})^T),$$
The matrix $\mathbf{C}_{Med}$ constitutes a median-based counterpart to the classical covariance matrix $\boldsymbol{\Sigma}$. The higher vector moments can be defined as
$$\boldsymbol{\Phi}_{b+1} = \sum_{a=0}^{b-1} (-1)^{a+1} \text{Med}((\mathbf{X} - \mathbf{M}) \mathbf{I}(\mathbf{X} \in S_a)), \quad b = 2, \dots$$
The third vector MedAD moments beyond location and scale is
$$\boldsymbol{\Phi}_3 = \text{Med}((X - \mathbf{M}) \mathbf{I}(X \in S_1)) - \text{Med}((X - \mathbf{M}) \mathbf{I}(X \in S_0)),$$
using two shells $S_0(0,1/2]$ and $S_1(1/2,1]$. This measure directional skewness, net asymmetry of the distribution and its $\|\boldsymbol{\Phi}_3\|$ is the strength of skewness. If the distribution is centrally symmetric, $\boldsymbol{\Phi}_3 = \mathbf{0}$. It is a robust vector measure of multivariate skewness identifying both the presence and direction of asymmetry. The fourth vector MedAD moments is
$$\boldsymbol{\Phi}_4 = -\text{Med}((X - \mathbf{M}) \mathbf{I}(X \in S_0)) + \text{Med}((X - \mathbf{M}) \mathbf{I}(X \in S_1)) - \text{Med}((X - \mathbf{M}) \mathbf{I}(X \in S_2))$$
Using three shells $S_0(0,1/3]$, $S_1(1/3,2/3]$ and $S_1(2/3,1]$. This measure directional peripheral dominance and separate centre versus peripheral behaviour and produce a vector peripheral descriptor and its norm $\|\boldsymbol{\Phi}_4\|$ measures strength of peripheral. A robust vector measure of directional tail dominance isolating asymmetry driven by extreme observations.

Note that the norm of these measures may be analogous to Mardia measures of skewness and kurtosis $\beta_{1,d}$, and $\beta_{2,d}$. Moreover, the standardized VMedAD vector moments is
$$\boldsymbol{\Psi}_k = \frac{\boldsymbol{\Phi}_k}{\|\boldsymbol{\Phi}_2\|}, \quad k = 2, \dots,$$
isolate shape independently of dispersion but without using covariance matrices. The standardized vector

$$\Psi_2 = \frac{\Phi_2}{\|\Phi_2\|},$$

captures pure directional imbalance, independent of scale and plays a role analogous to the standardized mean vector $\Sigma^{-1/2}E(X-\mu)$ in classical multivariate analysis, while avoiding reliance on finite second moments and covariance matrix. Under central symmetry $\Psi_2 = 0$.
The third higher order

$$\Psi_3 = \frac{\Phi_3}{\Phi_2}$$

is a scale-free vector of multivariate skewness, its direction shows where asymmetry occurs, and its norm ($\|\Phi_3\|$) measures the strength of skewness independently of dispersion like Mardia skewness (scalar). Under elliptical symmetry $\Psi_3 = 0$.
The fourth standardized moments

$$\Psi_4 = \frac{\Phi_4}{\Phi_2}$$

captures directional peripheral dominance and its norm ($\|\Phi_4\|$) measures the strength of peripheral independently of dispersion.

### 3.2 Sample VMedAD moments

Let the multivariate data $\{x_1, \ldots, x_n\} \subset \mathbb{R}^d$, be an i.i.d. sample from a multivariate distribution. All estimators below are finite-sample, median-based, and do not require moments to exist. Estimate the center **M** by

$$\widehat{\Phi}_1 = \widehat{M} = \arg \min_{m \in \mathbb{R}^d} \sum_{i=1}^n \|x_i - m\|$$

This choice ensures coherence with spatial depth and affine equivariance, the function spatial median in package "ICSNP" in R-software (Core Team, 2026) is used to compute it (Nordhausen et al., 2023). The central vectors can be computed for each observation as

$$u_i = x_i - \widehat{M}, \quad i = 1, \ldots, n.,$$

These centered vectors are the building blocks of all higher vector moments. The second VMedAD moment is estimated by

$$\widehat{\Phi}_2 = \text{Med}(u_1, \ldots, u_n)$$

The scalar dispersion estimator is $\widehat{\Phi}_2^{\text{Med}} = \text{Med}(\|u_i\|^2)^{1/2}$. The median covariance matrix as

$$\widehat{C}_{\text{Med}} = \text{Med}(u_i u_i^\top)$$

This matrix is robust, well-defined under heavy tails. The spatial depth estimator for each observation,

$$\widehat{D}(x_i) = 1 - \frac{1}{n}\sum_{j=1}^n \frac{x_j - x_i}{\|x_j - x_i\|}$$

the function depth spatial in package "ddalpha" in R-software (Core Team, 2026) is used to compute it (Pokotylo et al., 2019; Zuo & Serfling, 2000).
The second vector moment is estimated by

$$\widehat{\Phi}_2 = \text{Med}(u_1, \ldots, u_n)$$

For $b \geq 2$,

$$\widehat{\boldsymbol{\Phi}}_{b+1} = \sum_{a=0}^{b-1}(-1)^{a+1}\text{Med}(\boldsymbol{u}_i\, \mathbf{I}\{x_i \in S_a\})$$

The third vector moment (skewness) $\Phi_3$ Using two shells $\hat{S}_0 = (0, 1/2]$ and $\hat{S}_1 = (1/2, 1]$

$$\widehat{\boldsymbol{\Phi}}_3 = \text{Med}(\boldsymbol{u}_i\mathbf{I}_{\hat{S}_1}) - \text{Med}(\boldsymbol{u}_i\mathbf{I}_{\hat{S}_0})$$

Fourth vector moment (peripheral dominance) using three shells

$$\widehat{\boldsymbol{\Phi}}_4 = -\text{Med}(\boldsymbol{u}_i\mathbf{I}_{\hat{S}_0}) + \text{Med}(\boldsymbol{u}_i\mathbf{I}_{\hat{S}_1}) - \text{Med}(\boldsymbol{u}_i\mathbf{I}_{\hat{S}_2})$$

The fourth scaled moments is $\widehat{\boldsymbol{\Psi}}_k = \frac{\widehat{\boldsymbol{\Phi}}_k}{\widehat{\Phi}_2^{\text{Med}}}, k \geq 2$.

These are scale-free shape descriptors, analogous to covariance-standardized classical moments, but without requiring $\boldsymbol{\Sigma}$ or finite moments.

**Theorem 1.** Let $\boldsymbol{X} \in \mathbb{R}^d$ be a random vector with distribution $F$, and let $\{x_1, \ldots, x_n\}$ be an i.i.d. sample from $F$. Assume that
1. $F$ admits a unique multivariate median $\mathbf{M}$ (e.g., the spatial median),
2. The depth function $D(x; F)$ is continuous in $x$, and the induced depth distribution has no point masses,
3. For a fixed $b \geq 2$, the population shell-wise and median-wise are uniquely defined.

Then, for each fixed $b \geq 2$,

$$\widehat{\boldsymbol{\Phi}}_{b+1} \xrightarrow{P} \boldsymbol{\Phi}_{b+1},$$

where $\widehat{\boldsymbol{\Phi}}_{b+1}$ denotes the empirical vector MedAD moment estimator and $\boldsymbol{\Phi}_{b+1}$ its population counterpart. Moreover, the standardized vector moments

$$\widehat{\boldsymbol{\Psi}}_k = \frac{\widehat{\boldsymbol{\Phi}}_k}{\widehat{\Phi}_2^{\text{Med}}}, k \geq 2,$$

are consistent provided $\Phi_2^{\text{Med}} > 0$.

*Proof.* Consistency follows from the following facts. The spatial median estimator $\widehat{\mathbf{M}}$ is strongly consistent for M. Empirical depth functions and depth quantiles are consistent. This implying that the convergence of empirical depth shells to their population counterparts under the assumed continuity and absence of point masses in the depth distribution. Under weak convergence, coordinate-wise medians are continuous functionals. Since each VMedAD moment is a finite linear combination of shellwise medians, the continuous mapping theorem implies

$$\widehat{\boldsymbol{\Phi}}_{b+1} \xrightarrow{P} \boldsymbol{\Phi}_{b+1}.$$

Consistency of the standardized moments $\widehat{\boldsymbol{\Psi}}_k = \widehat{\boldsymbol{\Phi}}_k / \widehat{\Phi}_2^{\text{Med}}$ follows by Slutsky's theorem, provided $\Phi_2^{\text{Med}} > 0$.

**Theorem 2.** Assume the setup of Theorem 1. Then
1. The location estimator $\widehat{\boldsymbol{\Phi}}_1 = \widehat{\mathbf{M}}$ has a breakdown point of 50%,
2. The VMedAD scale estimator $\widehat{\Phi}_2^{\text{Med}}$ has a breakdown point of 50%,
3. For fixed $b \geq 2$, the VMedAD moment estimator $\widehat{\boldsymbol{\Phi}}_{b+1}$ has a finite-sample breakdown point

$$\varepsilon^*(\widehat{\boldsymbol{\Phi}}_{b+1}) \geq \frac{1}{2b}.$$

*Proof.* The spatial median $\widehat{\mathbf{M}}$ has a breakdown point of 50%, and hence so does the location estimator $\widehat{\boldsymbol{\Phi}}_1 = \widehat{\mathbf{M}}$. The VMedAD scale $\widehat{\Phi}_2^{\text{Med}}$, being defined as a median of squared radial distances, also inherits a 50% breakdown point. For fixed $b \geq 2$, the vector MedAD moment $\widehat{\boldsymbol{\Phi}}_{b+1}$ is constructed as a finite linear combination of shellwise medians, where each depth shell contains approximately a fraction $1/b$ of the sample. Since the median within each shell has a 50% breakdown point relative to that shell, contamination must exceed approximately $1/(2b)$ of the total sample to destabilize a shellwise median. Consequently, the finite-sample breakdown points of $\widehat{\boldsymbol{\Phi}}_{b+1}$ is of order $1/(2b)$ (Hampel et al., 1986; Donoho, 1982).

**Theorem 3.** Let $\boldsymbol{X} \in \mathbb{R}^d$ be a random vector with multivariate median $\mathbf{M}$, and let $\{x_1, \ldots, x_n\}$ be an i.i.d. sample from its distribution. Let $A$ be a non-singular $d \times d$ matrix and $b \in \mathbb{R}^d$, and define the affine transformation
$$Y = AX + c.$$
Denote by $\boldsymbol{\Phi}_b(X)$ and $\widehat{\boldsymbol{\Phi}}_b(X)$ the population and sample vector MedAD moments of order $b$, respectively, and by $\Phi_2^{\text{Med}}(X)$ and $\widehat{\Phi}_2^{\text{Med}}(X)$ the corresponding VMedAD scale. Then the following properties hold for all $b \geq 1$.

- Location equivariant

$$\boldsymbol{\Phi}_1(Y) = A\boldsymbol{\Phi}_1(X) + c, \widehat{\boldsymbol{\Phi}}_1(Y) = A\widehat{\boldsymbol{\Phi}}_1(X) + c.$$

- Vector moment equivariant

$$\boldsymbol{\Phi}_k(Y) = A\boldsymbol{\Phi}_b(X), \widehat{\boldsymbol{\Phi}}_b(Y) = A\widehat{\boldsymbol{\Phi}}_b(X), b \geq 2.$$

- Affine invariant of standardized moments

Let
$$\boldsymbol{\Psi}_b(X) = \frac{\boldsymbol{\Phi}_b(X)}{\Phi_2^{\text{Med}}(X)}, \widehat{\boldsymbol{\Psi}}_k(X) = \frac{\widehat{\boldsymbol{\Phi}}_b(X)}{\widehat{\Phi}_2^{\text{Med}}(X)}, b \geq 2.$$

Then

$$\boldsymbol{\Psi}_b(Y) = \boldsymbol{\Psi}_b(X), \widehat{\boldsymbol{\Psi}}_b(Y) = \widehat{\boldsymbol{\Psi}}_b(X), b \geq 2.$$

*Proof.* The multivariate median is affine equivariant, implying $\boldsymbol{\Phi}_1(Y) = A\boldsymbol{\Phi}_1(X) + b$ and $\widehat{\boldsymbol{\Phi}}_1(Y) = A\widehat{\boldsymbol{\Phi}}_1(X) + b$. Centered vectors therefore transform as $u_i(Y) = A\,u_i(X)$. Spatial depth is affine invariant, so depth shells are preserved under affine transformations. Since coordinate-wise medians commute with linear transformations and VMedAD moments are defined as alternating finite linear combinations of shellwise medians, vector equivariance $\boldsymbol{\Phi}_k(Y) = A\boldsymbol{\Phi}_k(X)$ holds for all $k \geq 2$, with an analogous result for the sample estimators.

## 4 Applications
### 4.1 Example
To illustrate the behavior of vector MedAD moments under asymmetric multivariate distributions, we consider a two-dimensional normal mixture model and MRSz $\gamma_2$ measure (Mori et al., 1994)

$$\boldsymbol{\gamma}_2 = E(\|Z\|^2 Z), \qquad Z = \Sigma^{-\frac{1}{2}}(X - \boldsymbol{\mu})$$

This measure produces vector skewness, and its norm measures the skewness strength but still required finite third moments, has zero breakdown point and is sensitive to outliers. A sample of size $n = 500$ is generated according to

$$X \sim 0.70\,\mathcal{N}_2((50,50)^{\text{T}}, 6^2 I_2) \;+\; 0.30\,\mathcal{N}_2((80,80)^{\text{T}}, 3^2 I_2),$$

where $I_2$ denotes the $2 \times 2$ identity matrix. This design produces a distribution that is asymmetric, due to the shifted minority component, non-elliptical, because of the mixture structure.

Figure 1 displays the scatter plot of the simulated observations. Two distinct clusters are visible. A dominant, diffuse cloud centered near $(50, 50)$, and a smaller, more concentrated cluster near $(80, 80)$. The multivariate spatial median lies within the main cloud. Superimposed on the plot are the vector skewness $\boldsymbol{\Phi}_3$, the vector peripheral dominance $\boldsymbol{\Phi}_4$ and $\boldsymbol{\gamma}_2$ measure. For $\boldsymbol{\gamma}_2$ the function "sampleSkew from "MultiStatM" in R-software (R Core Team, 2026) is used to compute it (Terdik & Toufer, 2026).

The vector $\boldsymbol{\Phi}_3$ and $\boldsymbol{\gamma}_2$ points from the median toward the smaller cluster, indicating directional skewness induced by the mixture imbalance. In contrast, $\boldsymbol{\Phi}_4$ captures a center–periphery contrast, reflecting the directional dominance of peripheral observations associated with the distant component.

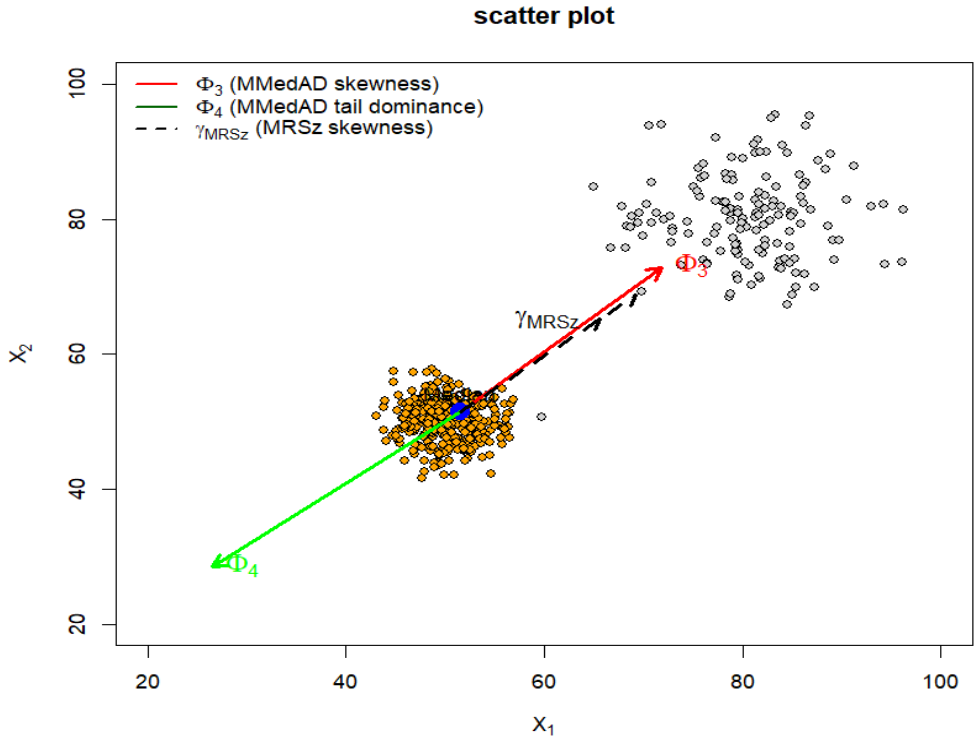

Figure 1. Scatter plot of a bivariate asymmetric normal mixture ($n = 500$), generated from $0.70 \, \mathcal{N}_2((50,50)^\mathrm{T}, 6^2 I_2) + 0.30 \, \mathcal{N}_2((80,80)^\mathrm{T}, 3^2 I_2)$.

## 4.2 Moments for an elliptical multivariate normal and $t$

Let multivariate elliptical normal $X \sim \mathcal{N}_d(\boldsymbol{\mu}, \boldsymbol{\Sigma})$, $\boldsymbol{\Sigma}$ positive definite with density

$$f(x) = |\boldsymbol{\Sigma}|^{-\frac{1}{2}} \exp((x - \boldsymbol{\mu})^\mathrm{T} \boldsymbol{\Sigma}^{-1}(x - \boldsymbol{\mu})/2)$$

The elliptical multivariate $t$ distribution extends the normal model by introducing heavier tails through an additional degrees-of-freedom parameter $\nu > 0$. A $d$-dimensional elliptical $t$ random vector can be represented as

$$X = \boldsymbol{\mu} + \boldsymbol{\Sigma}^{1/2} Z, \text{ where } Z = \frac{Y}{\sqrt{W/\nu}},$$

with $Y \sim N_d(0, I_d)$, $W \sim \chi^2_\nu$, and $Y$ independent of $W$.

Because VMedAD moments are affine equivariant, it suffices to compute them for the standardized case, and then transform back. Table 1 shows the results for normal and $t$ distributions.

Table 1. first four moments for an elliptical multivariate normal and t distributions

| Moments | $Z \sim \mathcal{N}_d(0, I_d)$ | $X \sim \mathcal{N}_d(\mu, \Sigma)$ | $Z \sim t_d(0, I_d, \nu)$ | $X \sim t_d(\mu, \Sigma, \nu)$ |
|---|---|---|---|---|
| $\Phi_1$ | 0 | M | 0 | M |
| $\Phi_2$ | 0 | 0 | 0 | 0 |
| $\|\Phi_2\|$ | $(F^{-1}_{\chi^2_d}(0.5))^{1/2}$ | $(F^{-1}_{\chi^2_d}(0.5))^{1/2}$ | $(\text{Med}(d F_{d,\nu}))^{1/2}$ | $(\text{Med}(d F_{d,\nu}))^{1/2}$ |
| $C_M$ | $\text{Med}(\chi^2_1) I_d$ | $\text{Med}(\chi^2_1) \Sigma$ | $\text{Med}(F_{1,\nu}) I_d$ | $\text{Med}(F_{1,\nu}) \Sigma$ |
| $\Phi_3$ | 0 | 0 | 0 | 0 |
| $\Phi_4$ | 0 | 0 | 0 | 0 |

The results for normal
- The distribution of $Z$ is centrally and spherically symmetric and rotation invariant, for a centrally symmetric distribution, the alternating sign construction guarantees, $\Phi_{2k+1} = 0, k = 1,2, ....$ Therefore, for the multivariate normal, $\Phi_3 = \Phi_5 = \Phi_7 = \cdots = 0$.
- The scalar scale is defined as $\Phi_2^{\text{Med}} = \text{Med}(\|Z\|^2)^{1/2}$. Since $\|Z\|^2 \sim \chi^2_d$, we obtain the exact population expression $\Phi_2^{\text{Med}} = (F^{-1}_{\chi^2_d}(0.5))^{1/2}$, which is the median of a chi-square distribution with $d$ degrees of freedom. For one population normal $\Phi_2^{\text{Med}} = 0.6745\sigma$,
- $C_{\text{Med}} = \text{Med}(UU^T)$. For $Z \sim \mathcal{N}_d(0, I_d)$: For $i \neq j$: $Z_i Z_j$ is symmetric about zero, median is zero. For $i = j$: $Z_i^2 \sim \chi^2_1$. Hence, $C_{\text{Med}} = \text{Med}(\chi^2_1) I_d$, with $\text{Med}(\chi^2_1) \approx 0.4549$.
- Even-order vector moments (e.g. $\Phi_4$) is rotational symmetry and depend only on radial shell contrasts. Consequently, their vector value is zero up to rotation. Only their norm carries information

For **an Elliptical Multivariate $t$-Distributio**
- Elliptical $t$-distributions are centrally symmetric, spherically symmetric, rotation invariant,
- An elliptical $t$ random vector admits the stochastic representation $X = \mu + \Sigma^{1/2} Z$, where $Z = \frac{Y}{\sqrt{W/\nu}}, Y \sim N_d(0, I_d), W \sim \chi^2_\nu$, with $Y$ and $W$ independent, the standardized case $Z \sim t_d(0, I_d, \nu)$,
- $\Phi_2^{\text{Med}} = \text{Med}(\|Z\|^2)^{\frac{1}{2}}$. For a $t_d(\nu)$ distribution, $\|Z\|^2 = \frac{\|Y\|^2}{W/\nu} \sim \frac{d F_{d,\nu}}{1}$, that is, a scaled $F$-distribution. Hence, $\Phi_2^{\text{Med}} = (\text{Med}(d F_{d,\nu}))^{1/2}$.
- For the standardized $t_d$ distribution: Off-diagonal elements $Z_i Z_j$ is symmetric about 0, $\text{Med}(Z_i Z_j) = 0, i \neq j$. Diagonal elements: $Z_i^2 = \frac{Y_i^2}{W/\nu}$, a scaled $F_{1,\nu}$ distribution. Then $C_{\text{Med}}(Z) = \text{Med}(F_{1,\nu}) I_d$
- Because the elliptical $t$ distribution is centrally symmetric, the alternating sign construction yields $\Phi_{2k+1}(Z) = 0, k = 1,2, ...$ In particular, $\Phi_3 = \Phi_5 = \cdots = 0$.
- Even-order vector MedAD moments (e.g. $\Phi_4$): encode center–periphery contrast, have no preferred direction due to spherical symmetry, depend only on the radial distribution. Therefore: the vector value is rotationally null, only their norms carry information.

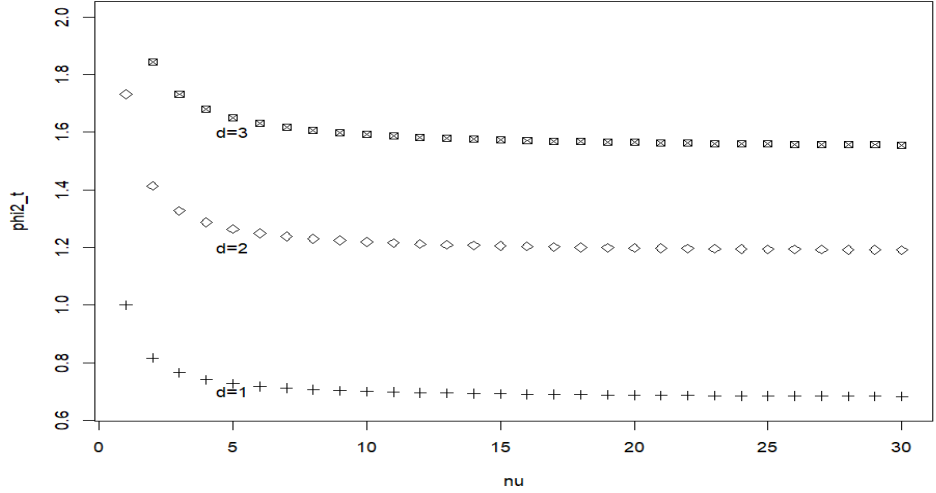

Figure 2. values for $\Phi_2^{Med}$ from multivariate $t$ distribution at $d = 1, 2$ and $3$

Figure 2 illustrates how the MedAD scale $\Phi_2^{Med}$ decreases with increasing degrees of freedom. For the multivariate $t$ distribution, converging to the corresponding Gaussian value, and thereby capturing tail heaviness in a robust, moment-free manner for $d = 1, 2$, and $3$. Unlike variance or covariance-based scales, the scale $\Phi_2^{Med}$ is well defined for the Cauchy distribution ($d = 1, v = 1$) which is equal to 1.

### 4.3 Wisconsin dataset

To demonstrate the practical relevance of VMedAD moments, we analyse the Wisconsin Breast Cancer (Diagnostic) dataset, consisting of 569 patients and 30 continuous tumour morphology variables measured from fine-needle aspirate biopsy images (Wolberg et al., 1995). This data is available in the site https://archive.ics.uci.edu/ml/machine-learning-databases/breast-cancer-wisconsin/wdbc.data.

For interpretability, the focus will be on two clinically meaningful variables, mean radius and mean concavity. These variables capture tumor size and shape irregularity; both strongly linked to malignancy. The data exhibit strong heterogeneity due to the presence of benign and malignant tumours, leading to asymmetric and heavy-tailed joint distributions. The VMedAD gives

$$\boldsymbol{\Phi}_1 = [13.36, 0.064], \boldsymbol{\Phi}_2 = [0.01, -0.002], \boldsymbol{\Phi}_3 = [1.775, 0.054], \boldsymbol{\Phi}_4 = [-1.525, -0.066,$$
$$\boldsymbol{\Psi}_3 = [0.934, 0.028], \quad \boldsymbol{\Psi}_4 = [-0.803, -0.035]$$

norms
$$\Phi_2^{Med} = 1.90, \|\boldsymbol{\Phi}_3\| = 1.776 \ \|\boldsymbol{\Phi}_4\| = 1.527, \|\boldsymbol{\Psi}_3\| = 0.934, \|\boldsymbol{\Psi}_4\| = 0.803$$

The Mardia
$$sk = 4.03, \quad ku = 14.984$$

The MRSz
$$\boldsymbol{\gamma}_1 = [1.046, 1.982], \quad \boldsymbol{ku}_1 = \begin{pmatrix} 1.21 & -0.37 \\ -0.37 & 5.76 \end{pmatrix}$$

The classical Mardia skewness and kurtosis statistics indicate substantial departure from multivariate normality; however, as scalar summaries, they do not reveal the geometric origin of this departure or identify which directions are driven by malignant extremes. MRSz points strongly in the mean radius direction, confirming that tumor size dominates the asymmetric structure provide directional information but aggregate contributions from all observations.

VMedAD moments yield a distinct geometric interpretation. The VMedAD skewness highlights the dominant direction associated with tumour malignancy, while the VMedAD peripheral dominance vector isolates the influence of extreme malignant cases in the outer depth shells. This decomposition demonstrates that much of the observed asymmetry arises from peripheral tumour morphology rather than from central benign structure, an insight not attainable using classical or projection-based multivariate moments.

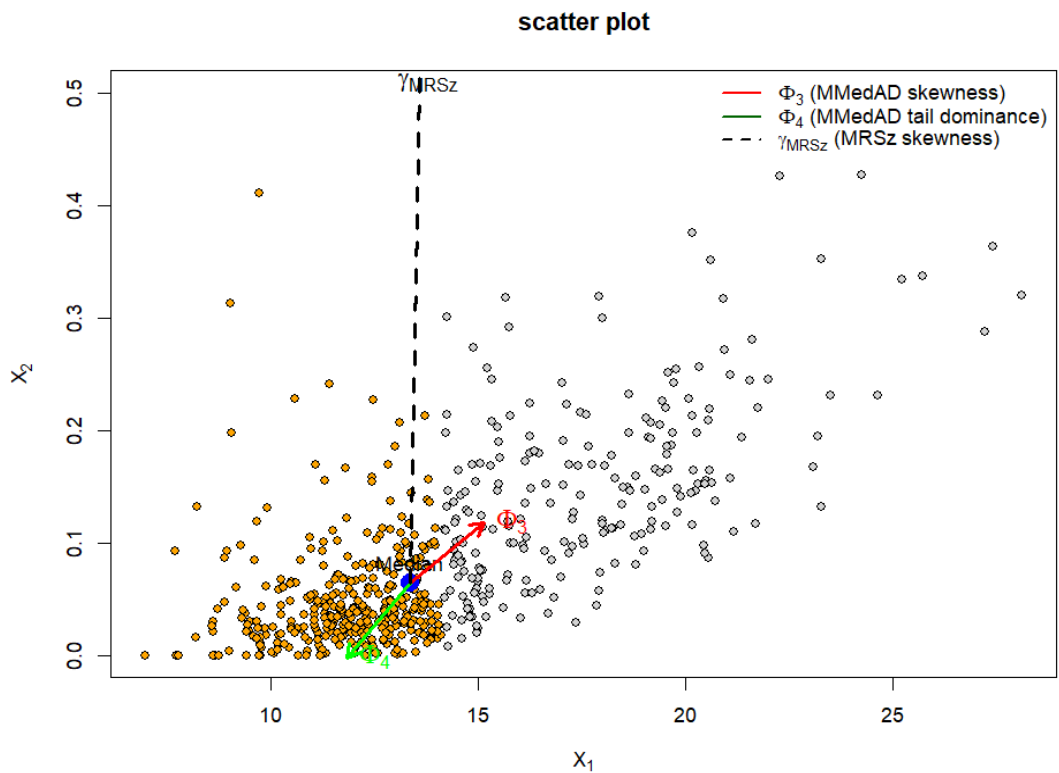

Figure 3. $\widehat{\boldsymbol{\Phi}}_3, \widehat{\boldsymbol{\Phi}}_4$ and $\boldsymbol{\gamma}_2$ for Wisconsin dataset

Figure 3 Scatter plot of mean radius ($X_1$) versus mean concavity ($X_2$) for the breast cancer Wisconsin (Diagnostic) dataset. The red arrow represents the VMedAD skewness $\widehat{\boldsymbol{\Phi}}_3$, indicating the dominant direction of multivariate asymmetry relative to the spatial median. The green arrow represents the VMedAD peripheral dominance $\widehat{\boldsymbol{\Phi}}_4$, isolating center–periphery contrast driven by extreme malignant cases. The dashed line corresponds to the MRSz skewness direction. While $\widehat{\boldsymbol{\Phi}}_3$ and MRSz identify similar skewness directions, $\widehat{\boldsymbol{\Phi}}_4$ reveals that the observed asymmetry is primarily driven by peripheral observations.

## 5   Conclusion

This paper introduced a new class of vector median absolute deviation (VMedAD) moments for multivariate shape analysis. This approach provides a robust and geometrically interpretable alternative to classical covariance-based moments. By replacing moment aggregation and covariance standardization with median-based centre–outward contrasts defined through data depth, the proposed framework yields affine-equivariant, moment-free vector measures of multivariate skewness and directional peripheral dominance.

VMedAD moments remain well defined under heavy-tailed distributions, including the Cauchy case, and explicitly separate central structure from tail-driven behaviour. Theoretical results established consistency, breakdown properties, and affine equivariant. The simulation from mixture normal and real dataset demonstrated improved robustness and interpretability compared with classical and projection-based multivariate skewness measures.

An additional advantage of the proposed framework is its natural extension beyond fourth-order behaviour. VMedAD moments are defined for arbitrary order $b \geq 2$ through depth-shell contrasts, enabling systematic exploration of higher-order multivariate shape features without requiring finite moments. This flexibility allows practitioners to tailor the level of shape resolution to the problem at hand while retaining robustness and geometric interpretability.

Despite these advantages, several limitations merit discussion. First, VMedAD moments depend on the choice of depth function; although spatial depth provides an intrinsic and affine-invariant construction, alternative depth notions may emphasize different geometric aspects of the data. Second, the current framework focuses on independent observations and static distributions; extensions to dependent data, or functional data remain open problems. Addressing these issues, along with developing formal inference procedures, asymptotic normality and computational refinements, constitutes an important direction for future research.